\documentclass[ %
prl,twocolumn,
superscriptaddress,
 amsmath,amssymb,
 aps,
floatfix,
]{revtex4-1}

\usepackage{graphicx}
\usepackage{epsfig,graphics,keyval}
\usepackage{amsfonts,epsf}
\usepackage{amsthm}
\usepackage{amsmath,bm}
\usepackage{amssymb}
\usepackage{psfrag}
\usepackage{color}
\def\u0#1{\underline {#1}}
\newcommand{\B}[1]{{\bm{#1}}}

\begin{document}

\title{Role of attractive forces in the relaxation dynamics of supercooled liquids}
\author{Joyjit Chattoraj}
\affiliation{School of Physical and Mathematical Sciences, Nanyang Technological University, Singapore}
\author{Massimo Pica Ciamarra}
\email{massimo@ntu.edu.sg}
\affiliation{School of Physical and Mathematical Sciences, Nanyang Technological University, Singapore}
\affiliation{CNR--SPIN, Dipartimento di Scienze Fisiche,
Universit\`a di Napoli Federico II, I-80126, Napoli, Italy}
\date{\today}

\begin{abstract}
The attractive tail of the intermolecular interaction affects very weakly the structural properties of liquids, while it affects dramatically their dynamical ones.
Via the numerical simulations of model systems not prone to crystallization, both in three and in two spatial dimensions, here we demonstrate that the non-perturbative dynamical effects of the attractive forces are tantamount to a rescaling of the activation energy by the glass transition temperature $T_g$:
systems only differing in their attractive interaction have the same structural and dynamical properties if compared at the same value of $T/T_g$.
\end{abstract}

\maketitle

According to the `van der Waals picture' the physics of liquids is dominated by the harsh and short-ranged repulsive forces between the particles, the weaker and longer ranged attractive forces only providing a homogeneous cohesive background.
This suggests the possibility of treating the attractive forces perturbatively, as first proposed by Weeks, Chandler and Andersen~\cite{WeeksCA1971,Chandler1983}. They indeed considered that, due the smooth spatial dependence of the attractive forces and the roughly homogeneous liquid structure, the sum of attractive forces experienced by a particle may be negligible with respect to the sum of the attractive ones. Hence `the arrangements and motions of molecules $\ldots$ are determined primarily by the local packing and steric effects produced by the repulsive forces'~\cite{Chandler1983}.
Berthier and Tarjus~\cite{BerthierTarjus2009,BerthierTarjus2011} investigated the validity of this scenario focusing on the Kob-Andersen binary Lennard-Jones (KA-LJ) model~\cite{KobAndersen1994}, a prototypical glass former. To asses the relevance of the attractive forces, they compared this model with its WCA variant (KA-WCA), where particles interact via the purely repulsive potential obtained by truncating the LJ potential at its minimum.
Their results demonstrated that the attractive forces have a non-perturbative effect on the relaxation dynamics, as the attractive forces greatly slow down the dynamics at low temperatures.
Subsequent works have clarified that the difference between attractive and purely repulsive interactions could be attributed to the small structural differences induced by the attractive forces. 
These differences have been first identified in higher order structural correlations~\cite{Coslovich2011,Coslovich2013,HockyMR2012,LiZS2016} and more recently, investigating two-point correlation functions via
machine learning techniques~\cite{LandesBDLR2019}. 
Relating the relaxation time to the configurational entropy through the Adam-Gibbs relation~\cite{Adam1965}, the effect of the different pair correlations on the dynamics has been rationalized considering their different contribution to the entropy~\cite{BanerjeeSSB2014, BanerjeeNSB2016, BanerjeeNSB2017}.
It has also been demonstrated that it is possible to design a purely-repulsive potential that seemingly generates many structural and dynamical properties of LJ liquids~\cite{PedersenSTD2010, BohlingVIBHTSD2012, DellSchweizer2015}.

All of these results indicate that in molecular liquids the attractive forces do not have a perturbative effect as originally speculated.
The open question ahead is therefore quantitatively rationalizing their non-perturbative influence.
In this Letter, we consider this question via the numerical investigation of the relaxation dynamics of a family of potentials characterized by the same repulsive part but different attractive tails. These potentials have been previously introduced to investigate the influence of the attractive interaction on the mechanical features of amorphous solids~\cite{DauchotKPZ2011}.
Our results demonstrate that attractive forces do not qualitatively change the features of the dynamics in the supercooled regime, but only rescale the typical temperature scale:
the structural and the dynamical properties of the different potentials coincide if their temperature is measured in units of their respective glass transition temperatures, $T_g$.
The non-perturbative role of the attractive forces is thus simply rationalized, to a very good approximation, via a simple rescaling of the typical energy scale.

\begin{figure}[t!]
  \begin{center}
    \includegraphics[width=0.47\textwidth]{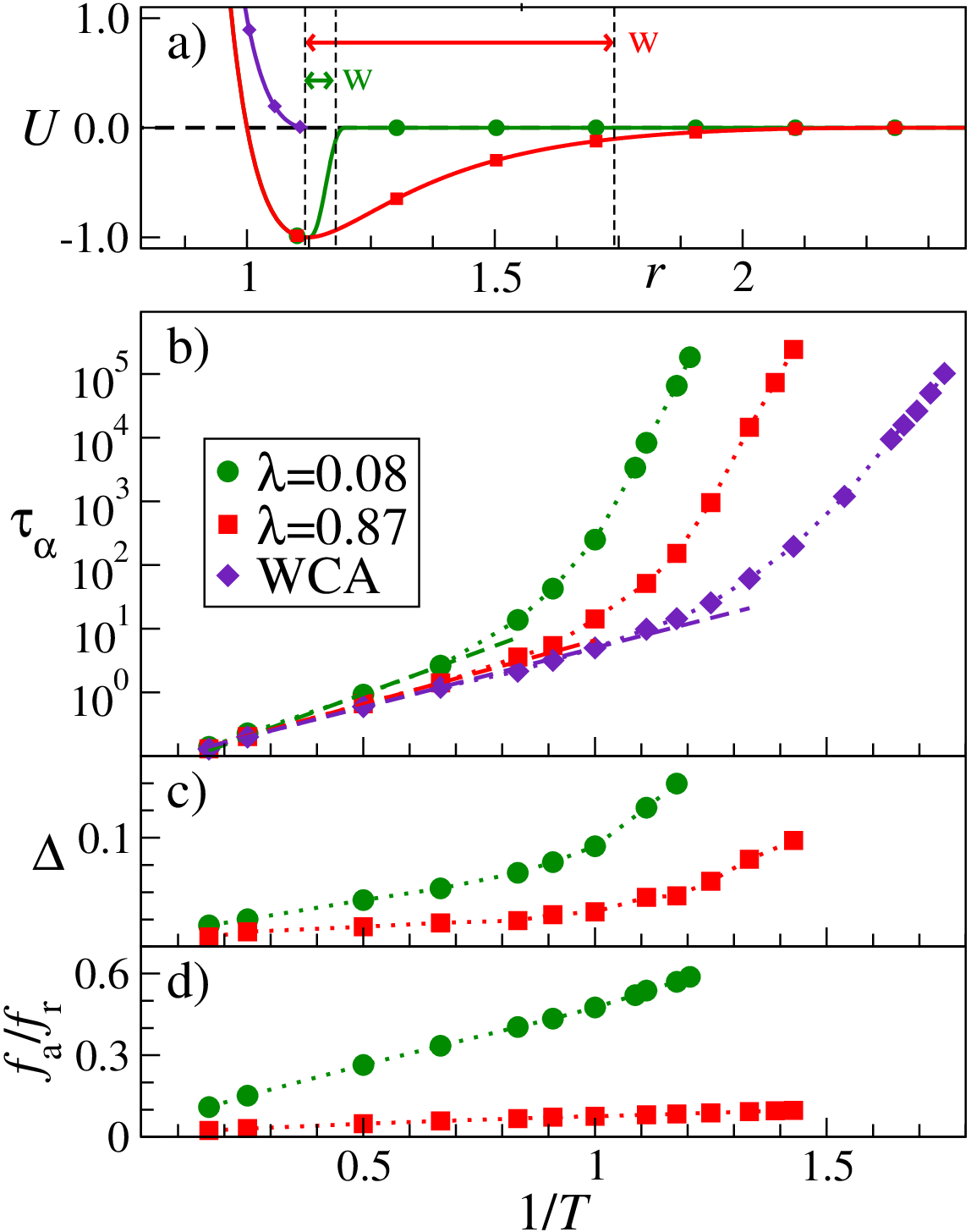}
 \caption{ 
 (a) The interaction potentials used in this study, and (b) dependence of their relaxation time on the inverse temperature. In (b), dashed lines are Arrhenius fits describing the high-temperature relaxation. 
 (c) accumulated deviation between two radial distribution functions $g(r)$ defined as $\Delta = \int^\infty_{0}\Delta(r;T)\mathrm{d}r$ where $\Delta(r;T)=|g_{\lambda}(r;T)-g_{\text WCA}(r;T)|$.
 (d) the ratio between the net attractive and the net repulsive force acting on a particle, averaged over all particles, increases upon supercooling.
}
\label{Fig:taualpha}
\end{center}
\end{figure}

{\it Model --}
We investigate the role of the attractive interactions via the numerical simulations of an interaction potential with a repulsive and an attractive part $U(r_{ij},\lambda) = U_r(r_{ij})+U_a(r_{ij},\lambda)$, the latter depending on a cutoff~\cite{DauchotKPZ2011}.
The repulsive component acts for $r_{ij} \leq r_{ij}^{\rm min} = 2^{1/6} \sigma_{ij}$, and is given by a LJ potential,
$U_r(r_{ij}) = 4\epsilon_{ij} \left[ (\sigma_{ij}/r_{ij})^{12} - (\sigma_{ij}/r_{ij})^{6} \right]$.
Conversely, the attractive component only acts for $r_{ij}^{\rm min} \leq r_{ij} \leq r_{ij}^{(c)}$, and is given by
\begin{equation}
    U_a(r_{ij},\lambda) = \epsilon_{ij} \left[a_0 \bigg(\frac{\sigma_{ij}}{r_{ij}}\bigg)^{12}\hspace{-0.4cm} - a_1 \bigg(\frac{\sigma_{ij}}{r_{ij}}\bigg)^{6}\hspace{-0.2cm} + \sum_{l=0}^3 c_{2l} \bigg(\frac{r_{ij}}{\sigma_{ij}}\bigg)^{2l} \right].
    \label{eq:pot}
\end{equation}
Here the six parameters $a_0,a_1$ and $c_{2l}$ are set such that $U(r_{ij})$ and its first two derivatives are continuous at the minimum, $r_{ij}^{\rm min}$, and at the cutoff, $r_{ij}^{(c)}$, where the potential vanishes.
As the cutoff is varied, the repulsive part of the potential remains unchanged, while the width of the attractive tail changes. 
We define the width of the attractive tail as $w(r^{(c)})=r^{\rm 10\%}-r^{\rm min}$, where $U(r^{10\%}) = 0.1 U(r^{\rm min})$, as illustrated in Fig.~\ref{Fig:taualpha}a, and compare each potential to the Lennard-Jones one via the parameter $\lambda = w(r^{(c)})/w_{\rm LJ}$.
For $\lambda = 1$ we recover the LJ potential, while in the limit $\lambda \to 0$ the attraction range vanishes and the potential becomes discontinuous at its minimum. 
We have investigated the above potential for different values of $\lambda$. 
Here we focus on two extreme values, $\lambda = 0.08$ and $\lambda = 0.87$, the other values behaving in an analogous way. 
Beside investigating the above potentials, we also consider the WCA one, $U_{WCA}=U_r(r_{ij})+c_{ij}$, where the constant $c_{ij}$ is such that the potential vanishes continuously at $r_{ij}^{\rm min}$. The WCA potential, which is not formally obtained from our potential of Eq.~\ref{eq:pot}, would correspond to the $\lambda \to \infty$ limit.
{\color{black} Finally, in the Supplementary Material~\cite{SM} we present results suggesting that our results do not depend on the specific form of the attractive potential.}

To prevent crystallization and demixing~\cite{LeonforteBTWB2005} we study polydisperse systems with $\epsilon_{ij}=\epsilon$, $\sigma_{ij}=(\sigma_i+\sigma_j)/2$, and $\sigma_i$ is drawn from a uniform random distribution in the range [0.8:1.2].
We use $\sigma$, $\epsilon$ and the mass of the particles $m$  as units of length, energy and mass, respectively.
In the numerical simulations~\cite{Plimpton1995}, we first equilibrate the system at the desired value of the temperature $T$ and of the density $\rho$ via NVT simulations, and then perform production runs in the NVE ensemble. All measures are taken in thermal equilibrium, when the system displays no aging behavior. 
We focus on three dimensional (3d) systems with $N = 32000$ particles, but we will also show that our main result remains valid in two spatial dimensions (2d). 
We consider two values of the density, $\rho = 1.07$, and $\rho = 1.15$. 
We remark that at the smallest temperature which we investigate the  coexisting density for $\lambda = 0.87$ is $\rho_{\rm coex} \simeq 0.97$.
Hence, we are investigating density values close to the coexistence region, where the role of attractive forces is expected to be of great relevance.

{\it Results --}
We measure the structural relaxation time $\tau_\alpha$ from the decay of the  self-intermediate scattering function $F_s(\B q,t) = \frac{1}{N} \langle \sum_{j=1}^N e^{i {\B q}\cdot({\B r}_j(t)-{\B r}_j(0)) }\rangle$, with $q \simeq 7$ corresponding to the first peak position of the static structure factor. 
Specifically, we extract $\tau_\alpha$ by fitting $F_s$ curves with exponential function $\sim e^{-t/\tau_\alpha}$ in the interval of $F_s\in [e^{-1}\pm 0.1]$.
Fig.~\ref{Fig:taualpha}b illustrates the dependence of the relaxation time on the inverse temperature for $\lambda = 0.08$ and $\lambda = 0.87$, and for the WCA model. We have checked that $\lambda = 0.87$ is already in the $\lambda \to 1$ limit in the considered temperature range, meaning that the system behaves as a LJ ones.
In all cases, the relaxation time exhibits a crossover from an Arrhenius to a super-Arrhenius temperature dependence, as in fragile glass-formers.
At low temperature, the relaxation time of the WCA, $\lambda = \infty$, is smaller than that observed at $\lambda = 0.87$, which is smaller than that at $\lambda = 0.08$. 
Hence, as the attraction range decreases, the dynamics slows down.
This confirms previous results~\cite{BerthierTarjus2009, WangXu2014, LandesBDLR2019}.
We have also checked~\cite{Coslovich2011,Coslovich2013,BanerjeeSSB2014,HockyMR2012,LandesBDLR2019} that the differences in the relaxation dynamics occur together with structural differences, we quantify comparing two-point correlation functions through the parameter $\Delta = \int^\infty_{0}|g_{ \lambda}(r;T)-g_{\text WCA}(r;T)|\mathrm{d}r$~\cite{Ingebrigtsen2012,WangXu2014}.
Panel c demonstrates that, as the temperature decreases, the structure of attractive potentials increasingly deviates from that of the WCA, and that the deviation becomes larger with the shorter range of the attractive interaction.

This observed behavior is consistent with the failure of the perturbation picture of liquids~\cite{WeeksCA1971}, according to which the sum of the attractive forces experienced by a particle should be small and negligible with respect to the sum of the repulsive forces experienced by the same particle. 
To explicitly verify that this assumption fails we evaluate the average net attractive force acting on the particles
$f_{\rm a} = \langle | \B f_{\rm a}^{(i)} | \rangle$, where
${\B f}_{\rm a}^{(i)} = \sum_j {\B f}_{ij}(r_{ij})\theta(r_{ij}-r_{ij}^{\rm{min}})$ is the net attractive force acting on particle $i$, $\theta(\cdot)$ being the Heaviside step function; similarly, we evaluate the average value of the magnitude of the net repulsive force acting on the particles, $f_{\rm r}$.  For the attractive forces to act as a perturbation, one would need $f_{\rm a}/f_{\rm r} \ll 1$. 
Fig.~\ref{Fig:taualpha}d illustrates that $f_{\rm a}/f_{\rm r}$ increases as the temperature decreases, and that it also decreases as the attraction range increases. This is qualitatively as expected.
At $\lambda = 0.87$ the attractive forces may still account for 10\% of the repulsive ones: they are not small enough to be negligible.
Notice that, regardless of the attraction range, 
in the $T \to 0$ limit $f_{\rm a}/f_{\rm r} \to 1$, as the system reaches a state of mechanical equilibrium.
\begin{figure}[t!]
\begin{center}
\includegraphics[width=0.45\textwidth]{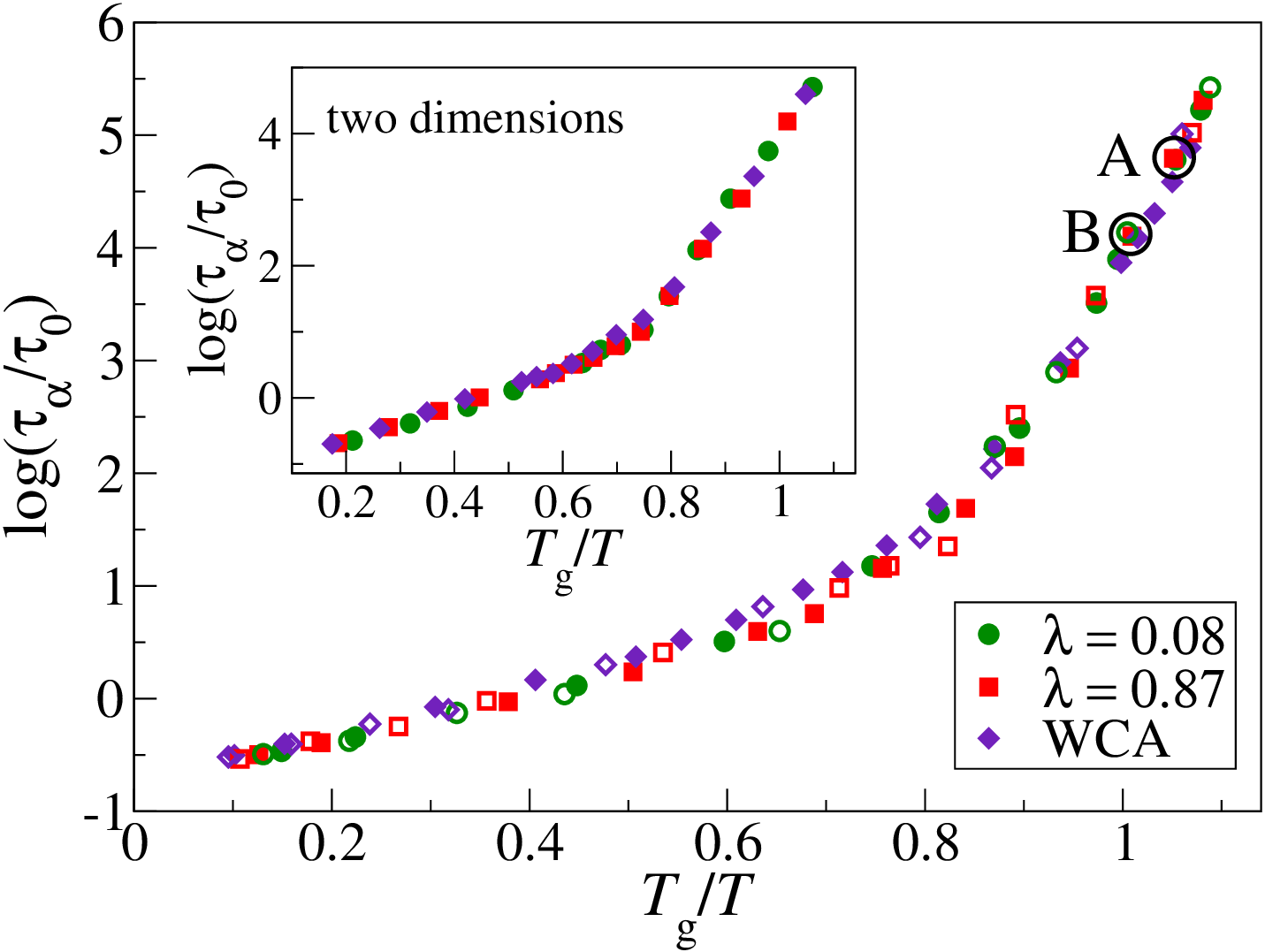}
\caption{ 
The main panel illustrates the dependence of $\log(\tau_\alpha/\tau_0)$ on $T_g/T$ for different potentials, as indicated. Full symbols are for $\rho = 1.07$, open ones for $\rho=1.15$.
{\textbf A} denotes a set of points ($\lambda=0.08, T=0.85, \rho=1.07$) and ($\lambda=0.87, T=0.72,\rho=1.07$) which have nearly same $\tau_\alpha$ and similarly, {\textbf B} denotes (WCA, $T=0.6,\rho=1.07$) and ($\lambda=0.87, T=0.75,\rho = 1.07$). 
The inset illustrates analogous two-dimensional results.
    }
    \label{Fig:Angell}
  \end{center}
\end{figure}

Having established that attractive forces play an important role in the relaxation dynamics, we now consider if they change the dynamics quantitatively, or also qualitatively. 
To this end, we investigate the Angell's plot~\cite{Angell1995} by operatively defining the glass transition temperature $T_g$ as that at which the relaxation time reaches $\tau_\alpha=10^4$. 
We find $T_g = 0.61$ for the WCA model, $T_g = 0.76$ for $\lambda = 0.87$ and $T_g = 0.9$ for $\lambda = 0.08$. 
Fig.~\ref{Fig:Angell} shows that when plotted versus $T_g/T$ the data of Fig.~\ref{Fig:taualpha}b collapse on a same master curve, at low temperature. Implying that all systems have the same fragility.
Notice that for dimensional consistency in Fig.~\ref{Fig:Angell} we have also rescaled the relaxation time by $\tau_0=1/\sqrt{T}$. 
We have verified that the collapse is robust and not affected by the definition of $T_g$. 
The inset of Fig.~\ref{Fig:taualpha}(b) shows that qualitatively analogous results hold in two dimensions. 
We have therefore reached the first important message of our investigation:
the non-perturbative effect of the attractive forces on the dynamics of supercooled liquids can be taken into account via a simple rescaling of the activation energy, for the model systems we have considered.
Note that the data collapse also holds to a good approximation at high temperature, where we observe an Arrhenius relaxation, 
$\tau_\alpha \propto \tau_0 \exp \left(\frac{\Delta E_{\rm A}}{T}\right)$.
This suggests $T_g \propto \Delta E_{\rm A}$. 
\begin{figure}[!t]
  \begin{center}
    \includegraphics[width=0.48\textwidth]{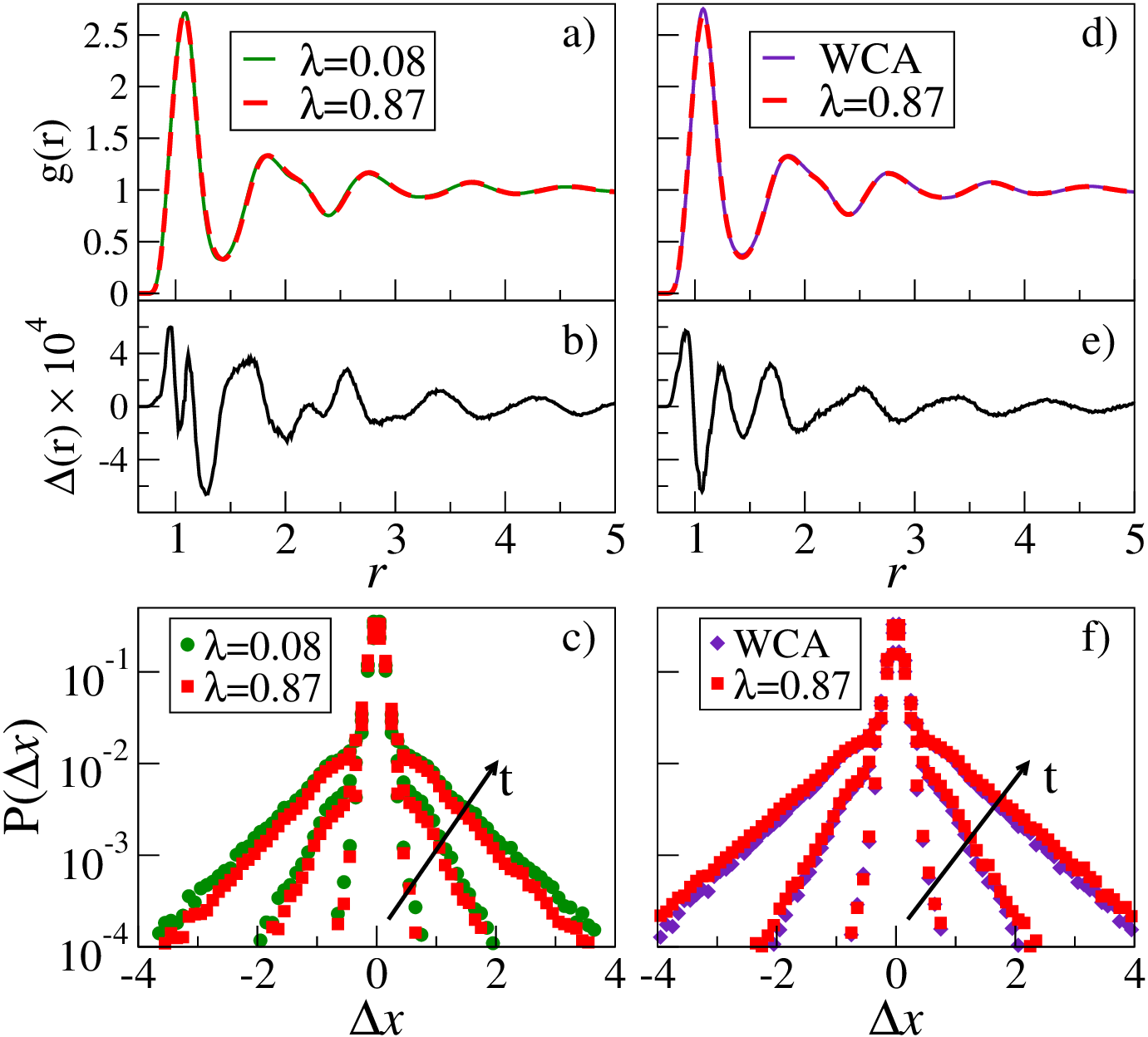}
    \caption{
      Panels a and b illustrate the radial distribution function for two different values of $\lambda$, and the radial dependence of their difference, i.e., $\Delta(r) = g_{\lambda^{(1)}}(r)-g_{\lambda^{(2)}}(r)$, for $T/T_g \simeq 0.94$. 
      Panel c compares the van-Hove distribution function of these systems, at times $t/\tau = 10, 10^3, 10^4$. 
      The same quantities are illustrated in panels d-f for the WCA potential and for $\lambda = 0.87$, at $T/T_g \simeq 0.98$.
      }
    \label{Fig:isomorphs}
  \end{center}
\end{figure}

This surprising result is related to a generalized isomorphysm induced by the attractive forces.
We remind~\cite{GnanSPBD2009, Dyre2013} that two state points of a same systems, having different densities and temperatures, are isomorph if they can be scaled into each other, meaning that their physical properties are identical when measured as a function of a thermodynamic parameter combining temperature and density. This exactly occurs in inverse power-law liquids, $U(r) \propto r^{-n}$, whose physical properties are fixed by $\rho^{n/d}/T$ in $d$ spatial dimensions.
To compare systems with different interaction potentials, but same density, here we postulate $T_g$ to be their relevant energy scale, and write the interaction energy as $U(r_{ij},\lambda) = T_g^{(\lambda)} u(\B r)$ with $u(\B r)$ a universal function. 
If this is so, then the statistical weight of configuration ${\B r}$,
\begin{equation}
    \exp \left(-\frac{U^{(\lambda)}(\B r)}{T} \right) = \exp \left(- \frac{u(\B r)}{T/T_g^{(\lambda)}} \right),
\end{equation}
does not depend on $\lambda$ if the temperature is measured in units of the glass transition temperature $T_g^{(\lambda)}$.
We validate this prediction in Fig.~\ref{Fig:isomorphs}. 
In panels a,b we compare the radial distribution functions at $T/T_g \simeq 0.94$, for $\lambda = 0.08$ and $\lambda = 0.87$.
In panel d,e we compare at $T/T_g \simeq 0.98$ the WCA potential, and the $\lambda = 0.87$ case. 
We observe the radial distribution functions to be indistinguishable, strongly supporting our speculation.
This generalized isomorphism also holds for the dynamical properties of the system. Fig.~\ref{Fig:Angell} already clarifies that when the temperature is measured in units of $T_g$ different potentials have the same relaxation time. 
In addition, we show in Fig.~\ref{Fig:isomorphs}c,f that state points at the same $T/T_g$ essentially share the same van Hove distribution of particle displacements, at all times. 
This implies that all of the dynamical properties of different systems are essentially identical, if not in the early ballistic regime.

The picture we have discussed so far remains valid as the density of the system increases. 
This is not surprising, as on increasing the density the role of the attractive forces become less and less relevant, so that the structural and the dynamical properties of the different systems converge.
This also occurs in KA-LJ systems~\cite{WangXu2014}. 
We have also checked that the picture remains true at lower density, as long as the investigated state points are not within the liquid-gas coexistence curve, for the attractive systems.

\begin{figure}[!!!t]
  \begin{center}
    \includegraphics[width=0.48\textwidth]{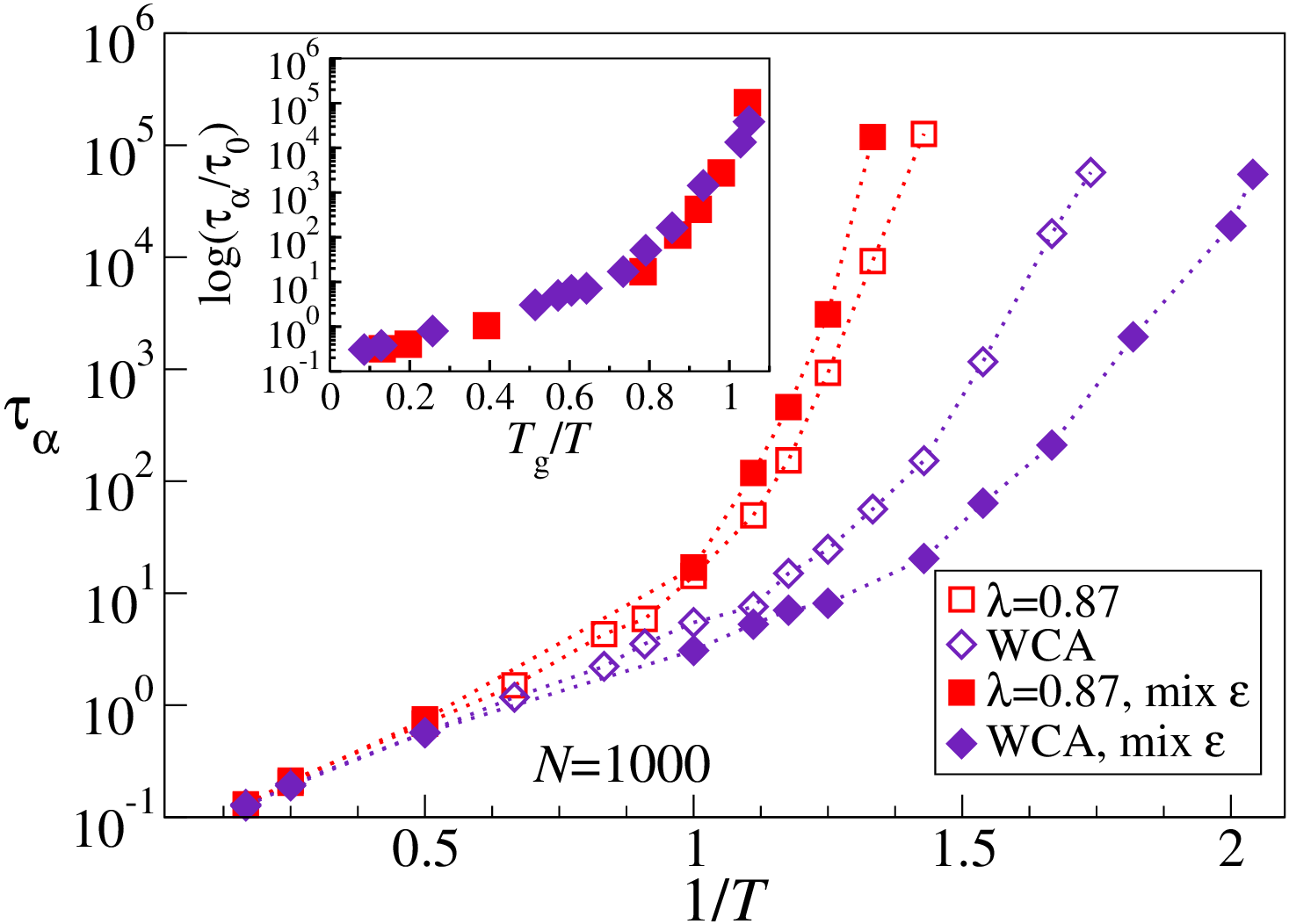}
    \caption{(a) Comparison of $\tau_\alpha$ between polydisperse systems (open symbols) and systems including both radial polydispersity
      and binary KA-type interaction strength (filled symbols).
      (inset) $\log(\tau_\alpha/\tau_0)$ as a function of $T_g/T$ for binary KA-type interaction $\epsilon$. 
    }
    \label{Fig:TauAlphaMixing}
  \end{center}
\end{figure}

Previous results have shown that rescaling the temperature via a typical energy is not enough to rationalize the difference in the relaxation dynamics of the KA-LJ and of the KA-WCA A-B 80-20 binary mixture~\cite{ElmatadChandlerGarrahan2010, BerthierTarjus2011}. 
In the KA-WCA model the fragility is density dependent, and in the KA-LJ it is again density independent~\cite{BerthierTarjus2009,BerthierTarjus2011}. 
On the contrary, we find the fragility to be potential and density independent.
We investigate the origin of this discrepancy considering that the KA model differs from our model in two aspects.
First, the KA model uses a binary rather than a continuous size polydispersity. 
Secondly, the mixing rule involves both the energy scales and the particle size, as $\epsilon_{AA} = \epsilon$, $\epsilon_{BB} = \epsilon/2$, $\epsilon_{AB} = 1.5 \epsilon$.
$\sigma_{AA} = \sigma$, $\sigma_{BB} = 0.88\sigma$, $\sigma_{AB} = 0.8\sigma$~\cite{KobAndersen1994, KobAndersen1995a}. 
To separately check the role of these two differences we perform simulations of an A-B 80-20 mixture using the energy mixing rule of the KA model, but a continuous polydispersity with $\sigma$ uniformly distributed in the [0.8:1.2] range. 
We fix the density to $\rho = 1.07$ and consider $80\%$ randomly chosen particles of type $A$, the other of type $B$. Fig.~\ref{Fig:TauAlphaMixing} shows the characteristic relaxation time over $T$ for these systems (full symbols). 
For comparison, the figure also presents data obtained without any mixing in energy, $\epsilon_{AA}=\epsilon_{BB}=\epsilon_{AB} = \epsilon$.
For $\lambda = 0.87$, energy mixing slightly slows down the dynamics. 
Conversely, for the WCA case it substantially speeds it up.
However, as illustrated in the inset of Fig.~\ref{Fig:TauAlphaMixing}, all data can be again collapsed by rescaling the temperature by the glass transition temperature. 
Hence, also in the presence of polydispersity in $\epsilon$ the attraction does not lead to qualitative changes in the relaxation dynamics, in the supercooled regime.
We therefore conclude that the fragility dependence~\cite{BerthierTarjus2009,BerthierTarjus2011} of the KA-WCA model is strongly affected by size bidispersity. 

In this respect, we have noticed that the bidispersity of KA-WCA is actually not enough to prevent its crystallization~\cite{Chattoraj_future}, and that it is actually easier to crystallize the KA-WCA than the KA-LJ, which also crystallizes~\cite{PedersenSD2018}. 
This suggests that the density dependence of the fragility of the KA-WCA~\cite{BerthierTarjus2009,BerthierTarjus2011} might be attributed to the presence of density and temperature dependent crystalline patches, or locally preferred structures~\cite{Coslovich2011}.

We have shown that the non-perturbative effect of the attractive forces on the dynamics of liquids can be taken into account via a rescaling of the relaxation energy scale: systems only differing in their attractive interaction have the same structural and dynamical properties at the same value of $T/T_g$, where $T_g$ is their respective glass transition temperature. 
Determining this energy scale from structural properties of the system remains an open issue.
We have checked that this picture holds in the liquid region of the phase diagram, also close to the coexistence curve. 

An important consequence of our finding is that the fragility is not affected by the attractive forces, but only by the repulsive ones which more directly control the local structure. This supports previous results which have linked the fragility to the emergence of locally favoured structures~\cite{Coslovich2007, RoyallWilliams2015, XiaLCKXFXW2015,WeiYJDYDDH2019} 
{\color{black} and also to the steepness of the repulsive potentials~\cite{WangGW2016}.} 
{\color{black} While we do have shown in the SM~\cite{SM} that our results are robust with respect to changes of the functional form of the attractive tails, we remain cautious about the generality of the role of attractive forces in the limit of very small attractive wells, where a re-entrant glass transition may occur~\cite{FDMST2005}.}
We finally remark that it has been shown that the glassy fragility is correlated with soft elastic modes~\cite{ShintaniTanaka2008, YanDW2013}. It is then interesting to understand how density of states behave for such polydisperse glasses where the fragility remains unaltered by attraction forces. We keep this question for future investigation.

\begin{acknowledgements}
 We thank Gilles Tarjus, Srikanth Sastry, Itamar Procaccia and Li Yanwei for useful discussions. 
 We acknowledge the support from the Singapore Ministry of Education through the Academic Research Fund (Tier 2) MOE2017-T2-1-066 (S), and from the NSCC for granting computational resources.
\end{acknowledgements}
\bibliographystyle{apsrev4-1}
%
 
\newpage
\setcounter{figure}{0}
\setcounter{equation}{0} 
\newcommand{\sFrac}[2]{{\textstyle\frac{#1}{#2}}}
\def\u0#1{\underline {#1}}
\def\theequation{S\arabic{equation}}
\renewcommand*{\thefigure}{S\arabic{figure}}

\onecolumngrid


\section{Supplemental Material - Effect of attraction with different functional form}
\begin{figure}[!b]
  \begin{center}
    \includegraphics[width=0.4\textwidth]{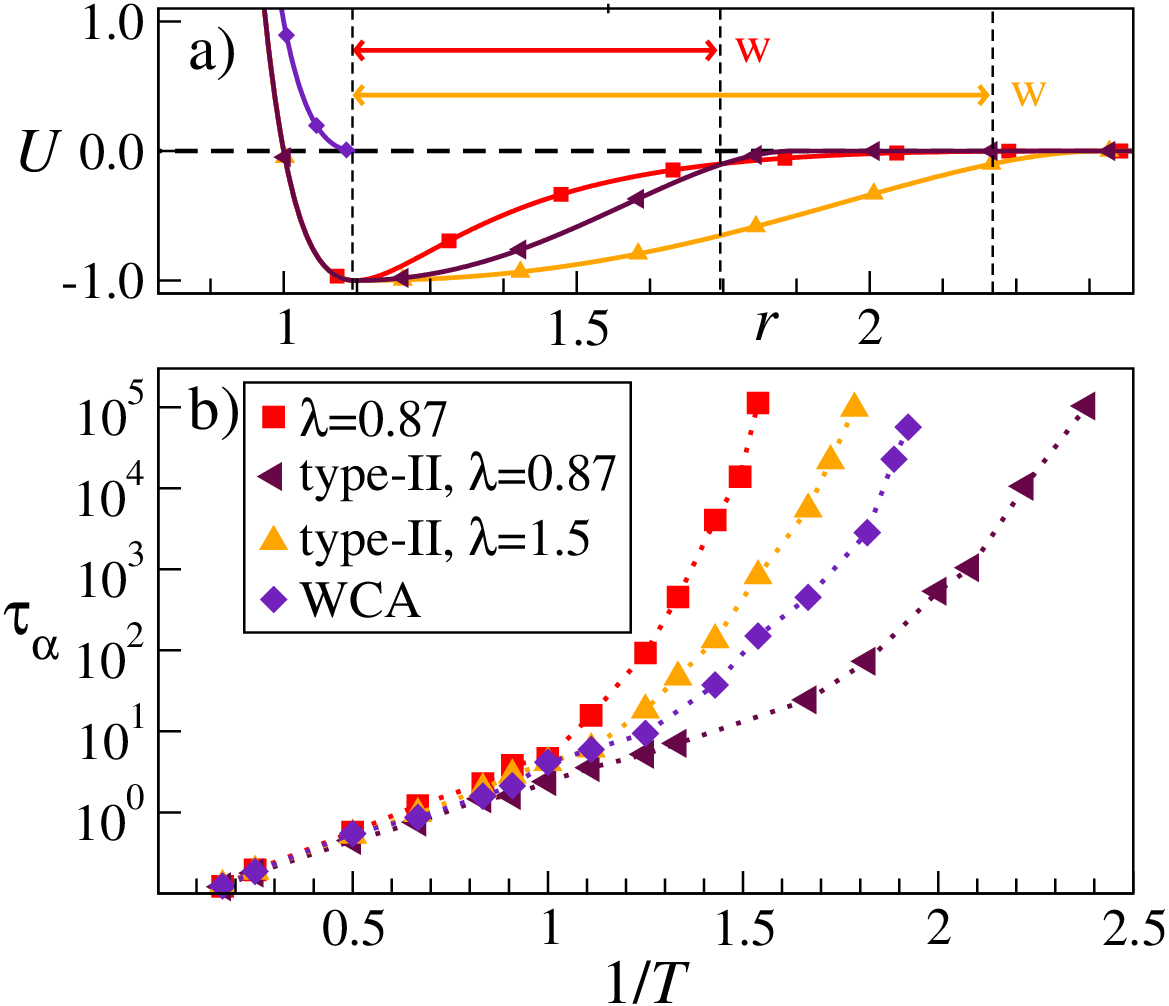}\hspace{2mm}
    \includegraphics[width=0.35\textwidth]{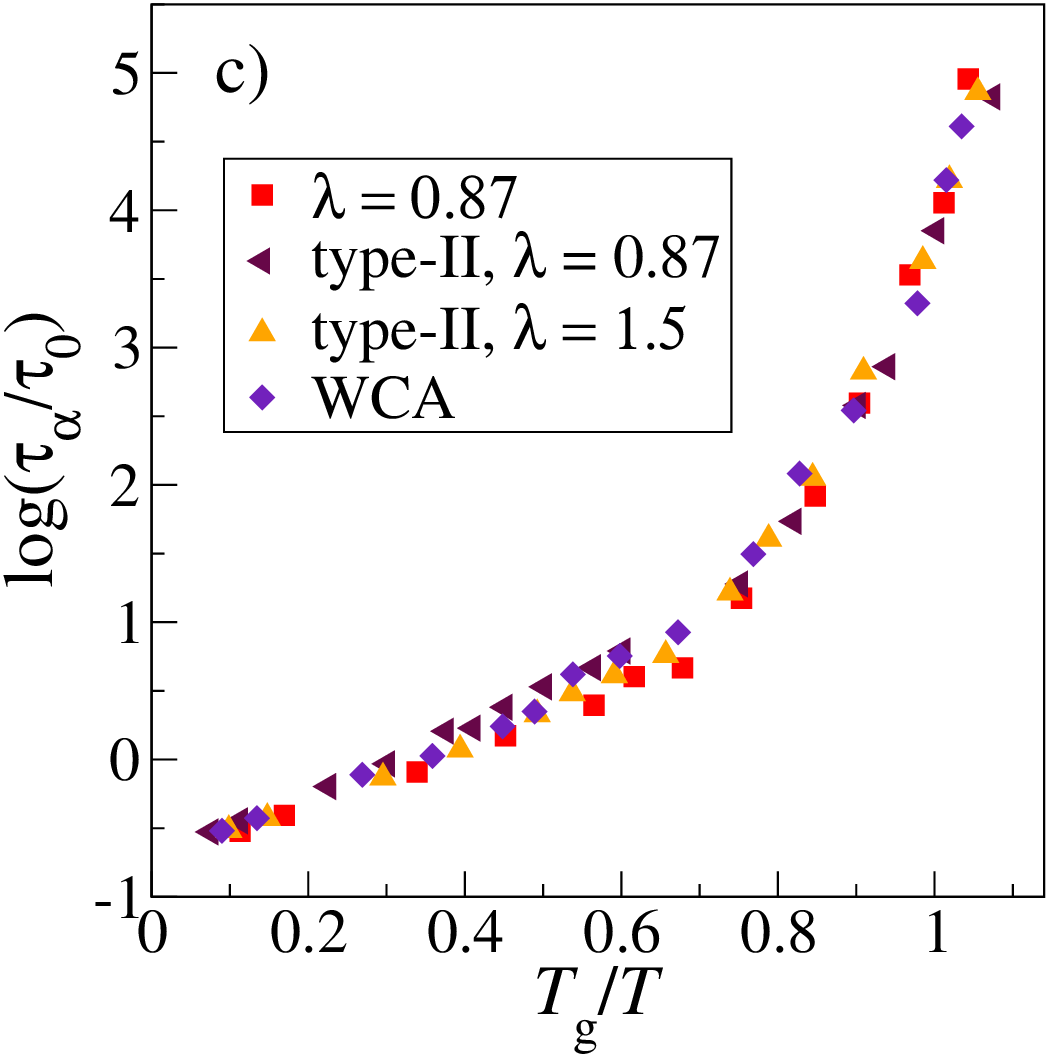}
 
    \caption{ (a) The functional form of the standard WCA and two different attractive potential interactions: (i) $U_a$ is a combination of Lennard-Jones 12-6 exponent terms and a polynomial term with $\lambda = 0.87$ (see Eq.~(1) in the main text), and (ii) $U_a$, referred type-II, is only a polynomial (see Eq.~(\ref{eq:potII})) with $\lambda = 0.87, 1.5$. (b) The structural relaxation time $\tau_{\alpha}$ for all the potential functions for a $N=1000$, $\rho=1.07$ sample. (c) $\tau_\alpha$ collapses after rescaling $T$ by $T_g$. 
    \label{Fig:tauPolyTail}
      }
  \end{center}
\end{figure}
We develop a potential function where the repulsive term follows the standard LJ potential, i.e.,
$U_r(r_{ij}) = 4\epsilon_{ij} \left[ (\sigma_{ij}/r_{ij})^{12} - (\sigma_{ij}/r_{ij})^{6} \right]$, when $r_{ij} \leq r_{ij}^{\rm min} = 2^{1/6} \sigma_{ij}$,
and the attractive term is expressed as a polynomial form
\begin{equation}
    U_a(r_{ij},\lambda) = \epsilon_{ij} \left[\sum_{l=0}^3 b_{2l} \bigg(\frac{r_{ij}}{\sigma_{ij}}\bigg)^{2l} \right] \ \ \textrm{for} \ r_{ij}^{\rm min} \leq r_{ij} \leq r_{ij}^{(\rm c)}.
    \label{eq:potII}
\end{equation}
The four coefficients $b_{2l}$ are set such that the potential function reaches its minimum value $-\epsilon_{ij}$ at $r_{ij}^{\rm min}$, and the function and its first derivative vanish at the cutoff $r_{ij}^{(\rm c)}$. We have  introduced the parameter $\lambda = w(r^{(c)})/w_{\rm LJ}$ to distinguish different attractive potential functions, where $w(r^{(c)})=r^{\rm 10\%}-r^{\rm min}$, is defined as the distance between the minimum position and the position at which energy is $0.1 U(r^{\rm min})$, and $w_{\rm LJ}$ follows the same definition for Lennard-Jones potential. We then study the structural relaxation time $\tau_\alpha$ over temperature $T$ for two different values of $\lambda=0.87, 1.5$ (see the two curves labelled type-II in Fig.~\ref{Fig:tauPolyTail}(b)). In comparison with the potential functions used in our main text, we find that with the new attraction potentials the relaxation time can go even faster than the relaxation time of WCA. Nevertheless after rescaling $T$ by the glass transition temperature $T_g$ ($\tau_\alpha \equiv 10^4$) we recover the same fragility for all potentials which we studied in this work (Fig.~\ref{Fig:tauPolyTail}(c)).    

\end{document}